\documentstyle[12pt,epsf]{article}
 \newcommand {\bi} {\bibitem}
 \newcommand {\be} {\begin{equation}}
\newcommand {\bea} {\begin{eqnarray} \nonumber }
\newcommand {\ee} {\end{equation}}
\newcommand {\eea} {\end{eqnarray}}
 \newcommand {\eps} {\epsilon}
 \newcommand {\si} {\sigma}
\newcommand {\de} {\delta}

\newcommand {\for} {\ \ \ \mbox{for}\ \ }

\newcommand {\cp} {\right)}
\newcommand {\ap} {\left(}

\def \form#1 {eq. (\ref{#1}) }
\def \parziale#1#2  {{\partial {#1} \over \partial {#2}}}

\begin{document}

\title{Numerical indications for the existence of a thermodynamic transition in binary
glasses}
\author{
Giorgio
Parisi\\
Dipartimento di Fisica, Universit\`a {\em La  Sapienza},\\ 
INFN Sezione di Roma I \\
Piazzale Aldo Moro, Rome 00185}
\maketitle

\begin{abstract}
In this note we present numerical simulations of binary mixtures and we find indications 
for a thermodynamic transition to a glassy phase.  We find that below the transition 
point the  off equilibrium correlation functions and response 
functions seems to be asymptotically compatible with the relations that were originally derived 
Cugliandolo Kurchan for generalized spin glasses.
\end{abstract}

\section{Introduction}

The behaviour of an Hamiltonian system (with dissipative dynamics) approaching equilibrium is well 
understood in a mean field approach for infinite range disordered systems \cite{CUKU,FM,BCKM}.  In 
this case we must distinguish an high and low temperature region.  In the low temperature phase the 
correlation and response functions satisfy some simple relations derived by Cugliandolo e Kurchan
\cite{CUKU}.  We present in this note the first investigation of the relations among these quantities 
for binary glasses.  We find indications for the existence of a phase transition.  The numerical 
data are compared to the theory.  The results of this comparison point toward the applicability of 
the results of the CK dynamical theory to binary glasses.

Generally speaking in a non equilibrium system it is natural to investigate  the 
properties of the correlation functions and of the response function.  Let us concentrate our 
attention on a quantity $A(t)$, which depends on the dynamical variables $x(t)$.  Later on we will 
make a precise choice of the function $A$.

Let us suppose that the system starts at time $t=0$ from a given initial condition and subsequently 
it follow the laws of the evolution at a given temperature $T$.  If the initial configuration is not 
at equilibrium at the temperature $T$, the system will display an off-equilibrium behaviour.  In many 
case the initial configuration is at equilibrium at a temperature $T'>T$; different results will be 
obtained as function of $T'$.  In this note we will consider only the case $T'>>T$ (in particular we 
will study the case $T'=\infty$).

We can define a correlation function
\be
C(t,t_{w}) \equiv <A(t_{w}) A(t+t_{w})>
\ee
and the response function
\be
G(t,t_{w}) \equiv \frac{ \de A(t+t_{w})}{\de \eps(t_{w})}{\Biggr |}_{\eps=0},
\ee
where we are considering the evolution in presence of a time dependent Hamiltonian in which we have
added the term
\be
\int dt \eps(t) A(t).
\ee

The off-equilibrium fluctuation dissipation theorem of the CK theory states some properties of the 
correlation functions in the limit $t_{w}$ going to infinity.  The usual equilibrium fluctuation 
dissipation (FDT) relation tell us that
\be G(t)= - \beta \frac{\partial C(t)}{\partial dt}, \ee
where
\be
G(t)=\lim_{t_w \to \infty} G(t,t_w), \ \ C(t)=\lim_{t_w \to \infty} C(t,t_w).
\ee
In our notation the correlation and response functions which depend on two times are the 
off-equilibrium ones; those which depend on only one time are the equilibrium ones, i.e.they are the 
limit of the off-equilibrium correlations when both times goes to at infinity at fixed distance.

It is convenient to define the integrated response:
\bea
R(t,t_{w})=\int_{0}^{t} d\tau G(\tau,t_{w}),\\
R(t)=\lim_{t_w \to \infty} R(t,t_w),
\eea
which is the response of the system to a field acting for a time $t$.

 We can also define the quantities
\be
C(\infty) = \lim_{t \to \infty} C(t), \ \ R(\infty) = \lim_{t \to \infty} R(t). 
\ee
The static FDT relation is 
\be
R(\infty)=\beta\ap C(0)-C(\infty)\cp.
\ee
The last equation can be naively written also as
\be
R(\infty)=\beta\ap <A^{2}>-<A>^{2} \cp \label{LINEAR},
\ee
Here the brakets denote the usual equilibrium expectation value.  If there is only one equilibrium 
state (or two that have opposite values $<A>$ and differ by a symmetry), the previous formula, 
\form{LINEAR} , is correct, otherwise a more lengthy discussione is need.

A very interesting situation happens when the quantity $<A>$ is identically zero because of symmetry 
arguments in the high temperature.  It is quite possible that there is a spontaneous symmetry 
breaking: two or more states of the systems may be present and the expectation of $A$ in the 
appropriate state becomes different from zero.  A typical example of this situation (see the 
appendix for more details) is given by spin glasses \cite{mpv,parisibook2,Noi}, where the magnetic 
susceptibility can be written at zero magnetic field as $
\chi= {\beta} / {N}\sum_{i=1,N}(<\si_{i}^{2}>-<\si_{i}>^{2})=\beta(1-q)$.

In this case the following relation is valid in the high temperature phase
\be
R(\infty)=\beta C(0),\label{HIGHT}
\ee
where we recall that in our notations $C(0)=\lim_{t\to\infty}C(t,t)$.  

The breaking at low temperature of the relation eq.  (\ref{HIGHT}) is a signal of a phase 
transition.  We will denote by $T_{c}$ the temperature at which the previous relation breaks.  We 
can also introduce an order parameter defined by
\be
q_{A}\equiv 1 -\frac{\beta C_{D}}{R(\infty)},
\ee
where
\be
C_{D}=\lim_{t\to \infty}C(t,t).
\ee

We can get further information on the nature of the transition if we stay in the framework of 
the CK theory for the approach to equilibrium \cite{CUKU}.  In the study of off-equilibrium spin 
glasses systems Cugliandolo and Kurchan proposed that the response function and the correlation 
function satisfy the following relations:
\be 
G(t,t_w)\approx-\beta X \ap C(t,t_w) \cp \frac{\partial C(t,t_w)}{\partial t}.
\ee
The previous relation can be also written in the following form
\be
R(t,t_w)\approx \beta \int_{C(t,t_w)}^{C(0,t_{w})}X(C) dC.
\ee
The function $X(C)$ is system dependent and its form tell us many interesting information.  

If $C(\infty)\ne0 $, we must distinguish two regions:
\begin{itemize}
\item A short time region where $X=1$ (the so called FDT region) and $C>C(\infty)$.
\item  A large time region (usually $t=O(t_w)$ where 
$C<C(\infty)$ and $X<1$ (the aging region) \cite{B,POLI}.
\end{itemize}

In the simplest non trivial case, i.e.  one step replica symmetry breaking, the function 
$X(C)$ is piecewise constant, i.e.
\bea
X(C)= m \for C<C(\infty),\\
X(C)= 1 \for C>C(\infty) \label{ONESTEP}.
\eea

In all known cases in which one step replica symmetry holds, the quantity $m$ vanishes linear with 
the temperature at small temperature.  It often happens (but it is not compulsory) that $m=1$ at 
$T=T_{c}$.

We  notice that we must be quite careful in exchange limits in the low temperature phase: the 
correlation function $C$ satisfy the relation
\be
\lim_{t\to\infty}C(t,t_w)=0
\ne\lim_{t\to\infty} \ap \lim_{t_{w}\to\infty}C(t,t_w) \cp=C(\infty).
\ee
In the sane way we have that in the region where $t$ and $t_{w}$ are {\it both } large
\bea
R(t,t_w)\approx R(\infty)=\beta(C(0)-C(\infty)) \for t<< t_{w}, \\
R(t,t_w)\approx R_{eq} = \beta \int_{0}^{C}dC X(C) \for t>> t_{w}.
\eea
Therefore it is quite possible that
\be
\lim_{t\to\infty}R(t,t_{w})\equiv R_{eq}\ne R(\infty).
\ee
This phenomenon is present as soon the function $X(C)$, is non zero outside the FDT region and it is 
the typical situation that happens when replica symmetry is broken \cite{mpv,parisibook2} .

The previous considerations are quite general.  However the function $X(C)$ is system dependent 
and its form tell us many interesting information.  Systems in which the replica symmetry is not 
broken are characterized by having $m=0$ in the formula \form{ONESTEP} .

Sometime simple aging is also assumed \cite{B}, i.e.  the following the scaling relation 
holds outside the FDT region:
\be
C(t,t_w)=C_{s}\ap{t \over t_{w}}\cp.
\ee

Simple aging may be correct, but it is not a necessary consequence of the previous relations and 
its verification is not the primary aim of  this note (a discussion of simple aging in the same 
system can be found in \cite{PAAGE1}).

The aim of this note is to show that binary mixture of spheres do have a transition in the 
thermodynamics sense at a temperature near the glassy transition, which can be characterized by a 
non zero value of the appropriate order parameter $q_{A}$.  More precisely we will show that there 
are equilibrium quantities which have an irregular (i.e.  non-analytic behaviour) at the transition 
point $T_{c}$.  Moreover the correlation and response functions seem to satisfy the relations of the 
CK theory, with the function $X(C)$ is compatible to be given by the one step formula 
(\ref{ONESTEP}).

The paper is organized as follows.  In section II, we define the model, the relevant quantities (i.e.  
the asymmetry or stress) and we present some general considerations.  In section III we study the 
approach to equilibrium of quantities defined at given time, e.g the energy and the 
equal time fluctuations of the stress.  We show that the fluctuations of the stress are strongly 
indicative of a phase transition.  In section IV we make a comparison of our data with the CK theory 
of aging.  Finally in the last section we present our conclusion.  At the end of the paper there is 
a short appendix where some results on spin glasses are recalled and a comparison is done with the 
finding of this paper.
 
\section{The model}
\subsection{The Hamiltonian}
The model we consider is 
the following.  We have taken a mixture of soft particles of different sizes.  Half of the particles 
are of type $A$, half of type $B$ and the interaction among the particle is given by the 
Hamiltonian:
\begin{equation}
H=\sum_{{i<k}} \left(\frac{(\si(i)+\si(k)}{|{\bf x}_{i}-{\bf x}_{k}|}\right)^{12},\label{HAMI}
\label{HAMILTONIAN}
\end{equation}
where the radius ($\si$) depends on the type of particles.  This model has been 
carefully studied  in the past \cite{HANSEN1,HANSEN2,HANSEN3,LAPA}.  It is known that a 
choice of the radius such that $\si_{B}/\si_{A}=1.2$ strongly inhibits crystallisation and the 
systems goes into a glassy phase when it is cooled.  Using the same conventions of the previous 
investigators we consider particles of average diameter $1$, more precisely we set
\begin{equation} 
{\si_{A}^{3}+ 2 (\si_{A}+\si_{B})^{3}+\si_{B}^{3}\over 4}=1.
\label{RAGGI}
\end{equation}
 
Due to the simple scaling behaviour of the potential, the thermodynamic quantities depend only on 
the quantity $T^{4}/ \rho$, $T$ and $\rho$ being respectively the temperature and the density.  For 
definiteness we have taken $\rho=1$.  The model as been widely studied especially for this choice of 
the parameters.  It is usual to introduce the quantity $\Gamma \equiv \beta^{4}$.  The glass 
transition is known to happen around $\Gamma=1.45$ \cite{HANSEN2}.

Our simulation are done using a Monte Carlo algorithm, which is more easy to deal with than 
molecular dynamics, if we change the temperature in an abrupt way.  Each particle is shifted by a 
random amount at each step, and the size of the shift is fixed by the condition that the average 
acceptance rate of the proposal change is about .4.  Particles are placed in  a cubic box with 
periodic boundary conditions.  In our simulations we have considered a relatively small number of 
particles $N=18$, $N=34$ and $N=66$.  We start by placing the particles at random and we quench the 
system by putting it at final temperature (i.e.  infinite cooling rate).

\subsection{The stress}
The main quantity on which we will concentrate our attention is the asymmetry in the energy (or 
stress):
\be
A=\sum_{{i<k}} \frac{(\si(i)+\si(k))^{12}} {|{\bf x}_{i}-{\bf x}_{k}|^{14}}
\ap 2(x_{i}-x_{k})^{2}-(y_{i}-y_{k})^{2}-(z_{i}-z_{k})^{2}\cp
 =2T_{1,1}-T_{2,2}-T_{3,3}.
\ee
In other words $A$ is a combination of the diagonal components of the stress energy tensor.
If the particles are in a cubic symmetric box, we have that
\be
<A>=0.
\ee
If the box does not have a cubic symmetry the effect of the boundary disappears in the infinite 
volume limit (at fixed shape of the boundary) and we have that the stress density $a$ vanishes in 
this limit:
\be
\lim_{N\to\infty}{<A>\over N} \equiv a =0.
\ee 
 
What happens when we add a term $\epsilon A$ to the Hamiltonian is remarkable.  Let us consider the 
new Hamiltonian
\be
H+ 12 \eps A,\label{PERH}
\ee
where $H$ was the old Hamiltonian \form{HAMI} .

 It is convenient to consider the following Hamiltonian:
\be 
H_{\epsilon}=\sum_{{i<k}} \left(\frac{(\si(i)+\si(k)}{r_{i,k}{(\eps)}}\right)^{12},\label{NEWH}
\ee
where
\be
 r_{i,k}(\eps)^{2}=
(x_{i}-x_{k})^{2}(1+\eps)^{-4}+ \ap (y_{i}-y_{k})^{2}+(z_{i}-z_{k})^{2}\cp (1+\eps)^{2}.
\ee
It is evident that we can recover the original Hamiltonian by contracting (for positive $\eps$) the 
$x$ direction by a factor $(1+\eps)^{-2}$ and expanding the $y$ and $z$ directions by a factor 
$1+\eps$, keeping in this way the volume constant.

As far as the expectation values of  intensive quantities do not depend on the shape of the box,
we can compute the properties of the theory with $\epsilon\ne 0$ in terms of those at $\epsilon=0$.
For example one finds that the energy and stress density are given by
\be
e(\epsilon)= e(0) \ \ \ \ a(\epsilon)=\frac{36 e(0)}{5} \epsilon   +O(\epsilon^{2}).
\ee

At the order $\eps$ the Hamiltonian in \form{NEWH} coincide with the previous one \form{PERH} .

We thus arrive to the following conclusion, if we consider the response of the system to adding an 
asymmetric term in the Hamiltonian.
\begin{itemize}
\item The equilibrium response function $R_{eq}$ is exactly given by $\frac{36}{5} e(0) $ at all 
temperatures.
\item
If there is only one equilibrium state, and this happens in the high temperature phase we must have 
at $\epsilon=0$
\be
{<A^{2}> \over N} = \frac{3}{5\beta} e.
\ee
\item
If we define (at $\epsilon=0$)
\be
C(t,t_{w})=12 \beta{<A(t+t_{w}) A(t_{w})> \over N}.
\ee
(where the factor $12\beta$ has been added to simplify the FDT relation), we obtain that in the high 
temperature phase
\be
 \lim_{t\to\infty} C(t,t)\equiv C_{D}=\frac{36}{5} e.
\ee
\end{itemize}

It is convenient to define the quantity
\be
W(t)=\frac{5 C(t,t)}{36 e(t)}\label{W}
\ee 
and investigate its limit for large time ($W\equiv\lim_{t \to \infty}W(t)$).  In the same way as in 
spin glasses we can define a quantity $q_{A}$ by
\be
q_{A}= 1- \frac{36 e}{5  C_{D}}= 1-W^{-1} \label{QA}
\ee
In the next section we shall see that there is transition from $q=0$ in the high temperature phase 
to a non zero value of $q$ at low temperatures.
\section{On the transition}
\subsection{General considerations}
We have done simulations for various values of $N$ ranging from $N=18$ to $N=66$.  For $N=18$ and 
$N=34$ we have measured the correlations by using 1000 runs with different 
starting point; for $N=66$ we have only 250 samples.  The evolution was done using the Monte Carlo 
method, with an acceptance rate fixed around .4.  At the end of each Monte Carlo sweep all the 
particles are shifted of the same vector in order to keep the center of mass fixed
\cite{LAPA}.  This last step in introduced in order to avoid drifting of the center of mass and it 
would be not necessary in molecular dynamics if we start from a configuration at zero total 
momentum.  Most of the quantities that we measure (with only one exception, i.e.  the quantity $Q$) 
to be defined later
\form{LATER} are not affected by this shift.

\begin{figure}[htbp]
\epsfxsize=400pt\epsffile[22 206 549 549]{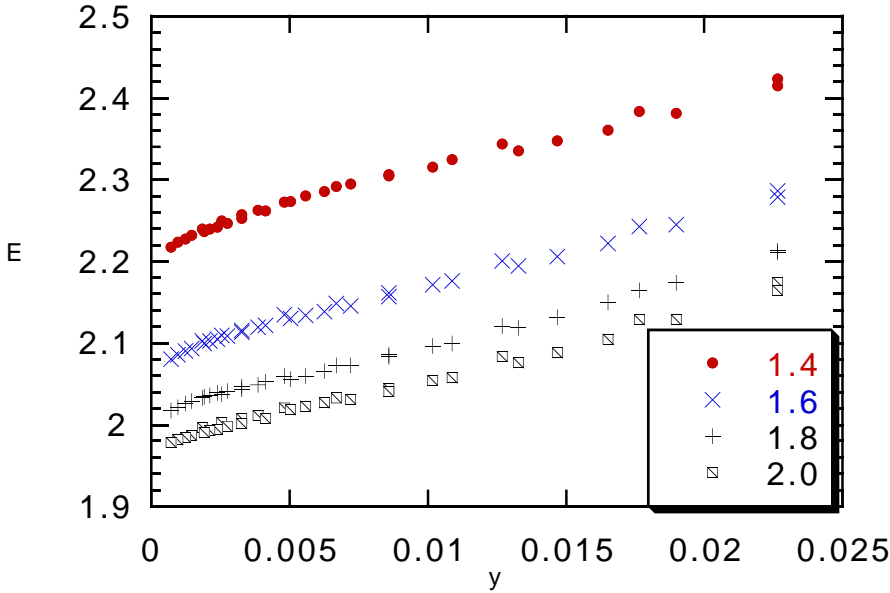}
\caption{The value of $E(t)$ versus $y\equiv t^{(-.7)}$ for different values of $\Gamma$ (1.4, 1.6, 
1.8 2.0) at $N=18$.}
\label{ENE18}
\end{figure}

\begin{figure}[htbp]
\epsfxsize=400pt\epsffile[22 206 549 549]{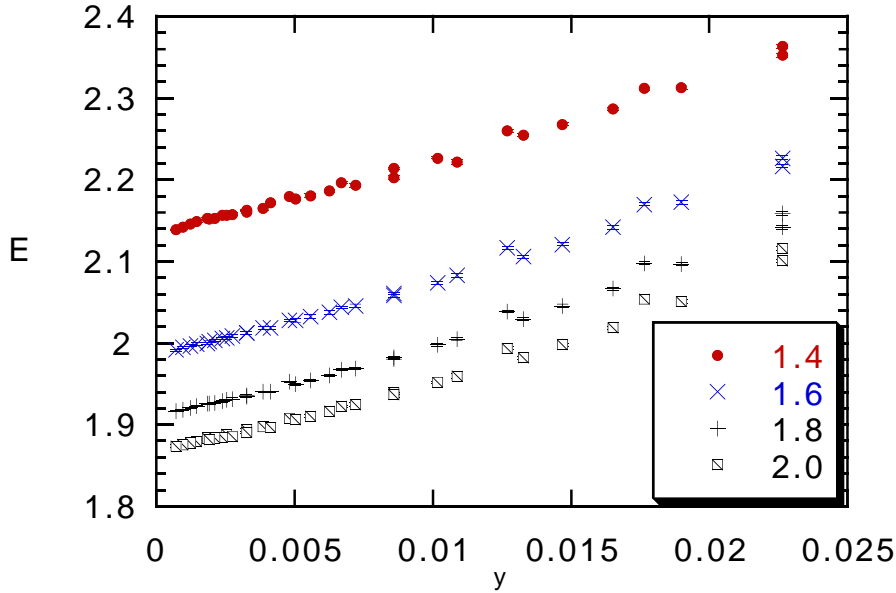}
\caption{The value of $E(t)$ versus $y\equiv t^{(-.7)}$ for different values of $\Gamma$ (1.4, 1.6, 
1.8 2.0) at $N=34$.}
\label{ENE34}
\end{figure}

Four observations are in order:
\begin{itemize}
\item
We need to do the average over many samples in order to decrease the error on the correlation 
$C(t,t)$.  Only one run would practically give no information because for this quantity the errors do 
not go to zero when $N$ goes to infinity.
\item
The values of $N$ may look rather small.  On the other hand our task is to show the existence of a 
thermodynamic phase transition.  Although definite conclusions can be only obtained by a careful 
finite size analysis for large $N$, strong indications of the existence of a transition can be 
obtained also for small samples.  For example in the Ising case the study of the susceptibility on a 
$4^{3}$ lattice is enough to draw strong indications for the existence of a transition and the 64 
binary degree of freedom are more or less the equivalent of our $54$ continuous degrees of freedom 
for $N=18$.
\item
The limit $t \to \infty $ may be tricky.  All the previous theoretical discussion were done for an 
infinite system.  In practice we need that that $N$ is greater that the time to a given power, the 
exponent of the power being system dependent.  Therefore we cannot strictly take the limit $t
\to \infty $ for finite $N$.  We have to study the behaviour of the system in region of time whose 
size increase with $N$.  Finite size corrections sometimes become much more severe at large times
\cite{BK, PAAGE1}.
\item
In this note we have explored a region of not very large times and we have extrapolated the data to 
infinity by assuming simple powers corrections at large times.  The real situation is likely  more 
complicated.  It is quite possible that when we fast cool the system we end up in a metastable state 
of energy higher the equilibrium one and that the system is going to relax to the equilibrium values 
with a much slow process, which cannot be seen on this time scale and produces a drift on a much 
slower scale.  In this situation our results concern not the real equilibrium value of the energy 
(and of the other quantities), but their value in a metastable state.
\end{itemize}
\subsection{The energy}
Let us consider the time dependence of the energy.  In fig.  (\ref{ENE18}) and (\ref{ENE34}) we see 
the energy for $\Gamma=1.4,1.6,1.8,2.0$, which are near or below the value $\Gamma=1.44$ at which 
the phase transition is estimated from the behaviour of the equilibrium correlation functions.

\begin{figure}[htbp]
\epsfxsize=400pt\epsffile[22 206 549 549]{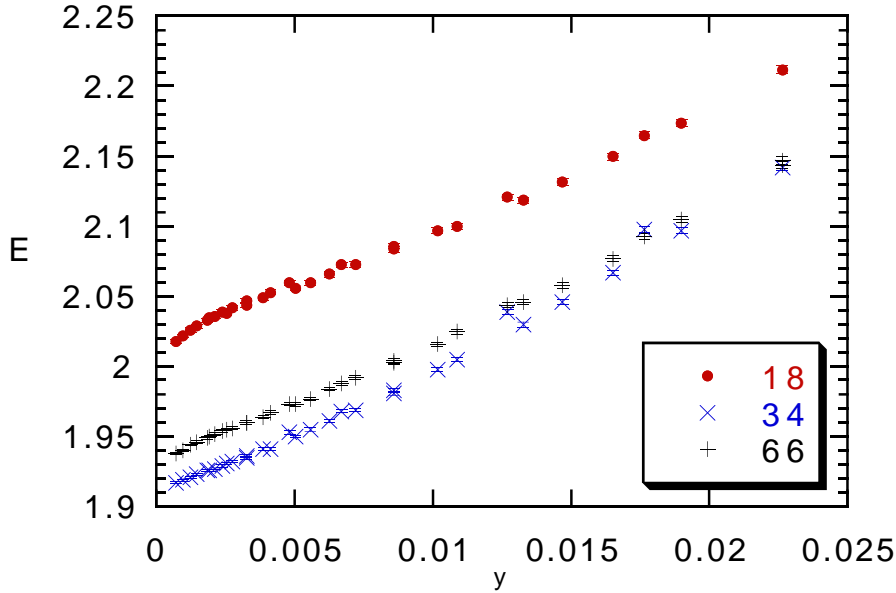}
\caption{The value of $E(t)$ versus $y\equiv t^{(-.7)}$ for different values of $N$ (18, 34 and 66) 
at $\Gamma=1.8$.  }
\label{ENETOT}
\end{figure}

The data are plotted as function of $y\equiv t^{(-.7)}$.  The data are quite linear as function of 
$y$ at $N=34$, while they show some curvature at large $t$, small $y$ for $N=18$.  This seems 
to be a small finite volume effect.  Apart from this effect and an overall shift, the two sets of 
data are quite similar so that we have reasons to suppose  that the $N=34$ are nearly 
asymptotic.  This is confirmed by a run at $N=66$ for $\Gamma=1.8$: a comparison of the three runs 
at $\Gamma=1.8$ is showed in fig.  (\ref{ENETOT}).

The results seem to be puzzling.  The energy goes to its asymptotic values with a power correction 
and the exponent of power correction is independent from the temperature when the temperature changes 
of about a factor 4 (and also it is not a small exponent!).  On the other hand we know that in real 
systems the convergence of the energy to the asymptotic value is extremely slow below the glass 
transition.  It is quite likely that we are blind to this slower process  that happens on a much 
slower time scale.  Moreover the temperature independence of the exponent strongly suggests that the 
process is not dominated by activated processes which become dominant at very larger times.

The situation is quite reminiscent of the mean 
field theory case, where the power to approach to a metastable state depends weakly on the 
temperature \cite{FRAMAPA}.  We can thus suppose that here also we converge to a metastable state, 
whose energy may be larger that the equilibrium one.  It would be interesting to verify this point 
by carefully computation of the true equilibrium energy by more tuned simulations.  We can thus 
tentatively conclude that our large time extrapolation do concern some kind of metastable state.  
If this happens, it is quite remarkable that the two time scales (controlling respectively the the 
approach to the metastable state and the slow decay of the metastable state) are so separated that 
the first can be studied independently from the second.

\subsection{The fluctuation of the stress}

Here we perform the same analysis as in the previous subsection for the quantity $W(t)$ defined in 
eq.  (\ref{W}), but we are very interested in finding the exact value of extrapolated at infinite 
time.  In the high temperature region we find that $W$ extrapolates to a value very near to 1, as 
expected from the previous considerations.

\begin{figure}[htbp]
\epsfxsize=400pt\epsffile[22 206 549 549]{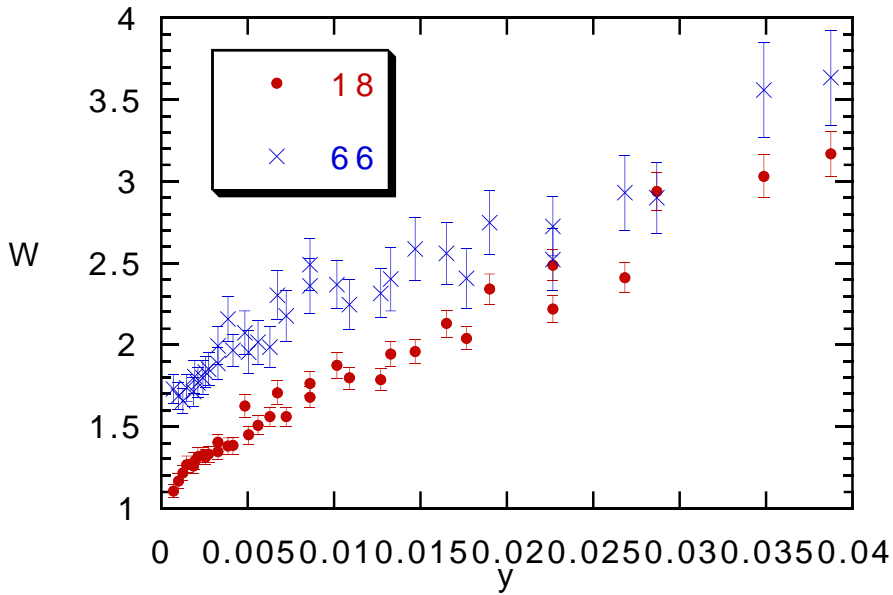}
\caption{The value of $W(t)$ versus $y\equiv t^{(-.7)}$ for $\Gamma=1.8$ for $N=18$ and  $N=66$.}
\label{PLOT18}
\end{figure}

New phenomena appear when we decrease the temperature below the one corresponding to $\Gamma=1.4$.  
In fig.  (\ref{PLOT18}) we show the data of $W(t)$ versus $y\equiv t^{(-.7)}$ for $\Gamma=1.8$ for 
$N=18$ and $N=66$.

Here also we the same phenomena as for the energy.  On the large sample the data are quite linear 
when plotted versus $y$.  It is remarkable that the same choice of power which (i.e.  .7) which 
works for the energy, is also good for $W$.  It seems that we have the same power corrections for 
both quantities.

The data at small lattice show a curvature in the plot 
versus $y$, which is absent in larger sample.  This is finite volume effect which indicates which is 
the range of time for which we can comfortable assume that the volume is sufficiently large.

\begin{figure}[htbp]
\epsfxsize=400pt\epsffile[22 206 549 549]{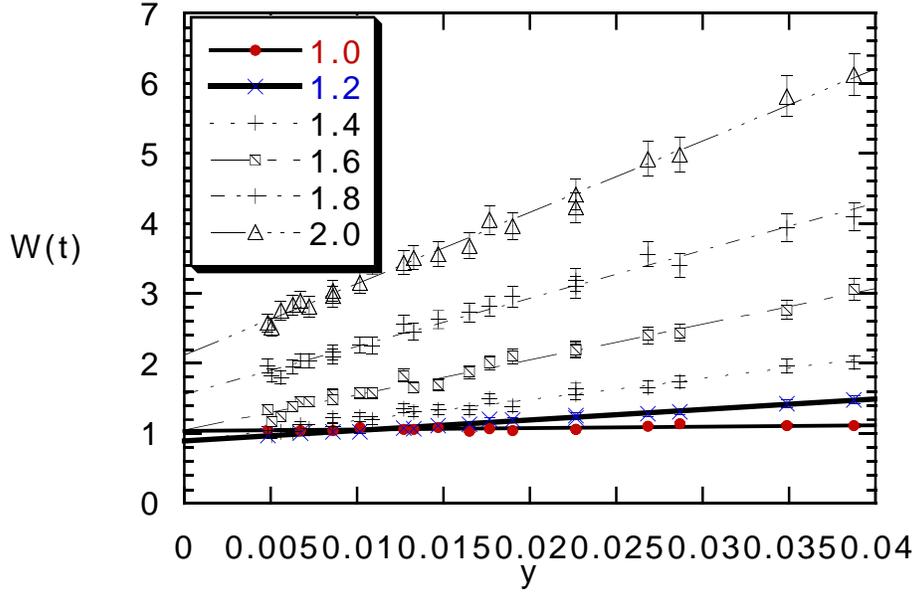}
\caption{The value of $W(t)$ versus $y\equiv t^{(-.7)}$ for different values of $\Gamma$ (1.0, 1.2, 
1.4, 1.6, 1.6 and 2) at $N=34$.}
\label{W34}
\end{figure}

\begin{figure}[htbp]
\epsfxsize=400pt\epsffile[22 206 549 549]{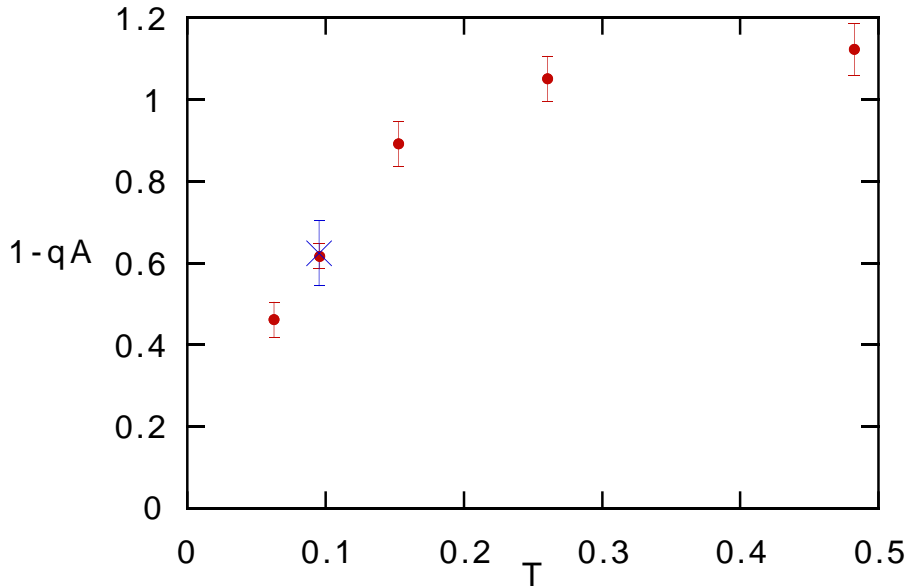}
\caption{The value of $1-q_{A}=W(\infty)^{-1}$ as function of the temperature using the data of $N=34$ 
extrapolated from the region of time $200-2000$, i.e.  the linear fits of the previous figures.  The 
cross at $\Gamma=1.8$ is the value extrapolated from the region of time $200-6400$ at $N=66$.}
\label{QGLASS}
\end{figure}

We show in fig.  (\ref{W34}) the value of $W(t)$ versus $y\equiv t^{(-.7)}$ for different values of 
$\Gamma$ at $N=34$.  A liner fit is rather good and the extrapolated value of $W(\infty)^{-1}$ are 
shown in fig. (\ref{QGLASS}).

It is clear from both fig.  (\ref{W34}) and (\ref{QGLASS}) that $W(\infty)$ becomes different from 1 
in the low temperature region.  If we assume that it diverges at low temperature, the data are 
compatible with the usual situation where  $W(\infty)$ is roughly proportional to the 
temperature at small temperature.  This results should be taken as an indication that, at least in 
the metastable region there is a transition to a region where $W(\infty)\ne 1$.

\section{The comparison with the CK Theory}
\subsection{The correlation functions}
Up to now we have considered equal time correlation functions.  Let us see what happen to the 
correlation functions at different times.  We will study the correlation $C(t,t_{w})$.

We introduce the variable $s=t/t_{w}$.  According to simple aging the correlation functions should 
become a function of only $s$ in the limit of large times.  Of course the FDT region, which is 
located at finite $t$ also when $t_{w}$ goes to infinity, is squeezed at $s=0$, so that we expect 
that the function $C$ becomes discontinuous at $s=0$.  Moreover the limit $s\to 0$ give us information 
on the value of $C(\infty)$
\be
\lim_{s \to 0}C(t,t_{w})=C(\infty),
\ee
where it is understood that the limit is done always remaining in the region where $t>>1$.

\begin{figure}[htbp]
\epsfxsize=400pt\epsffile[22 206 549 549]{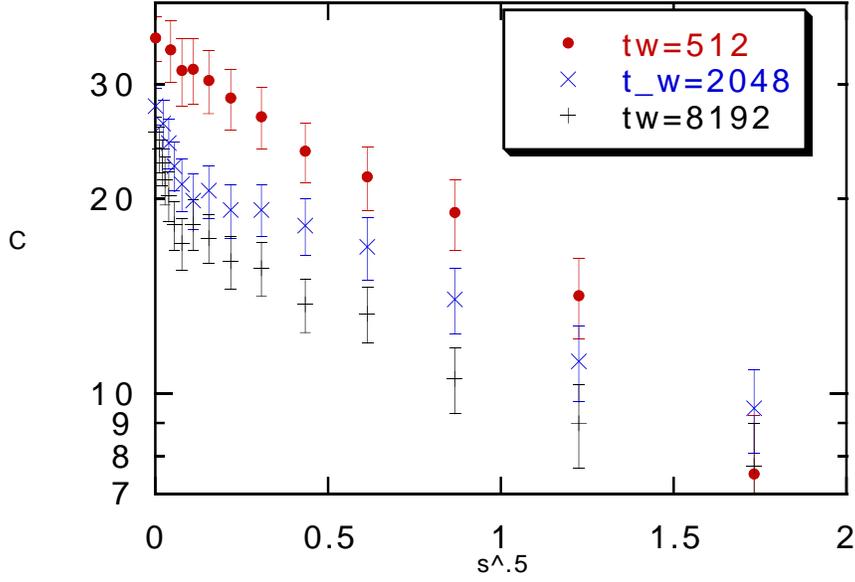}
\caption{On a logarithmic scale we see the correlation function at $\Gamma=1.8$ for $N=66$ at 
different values of $t_{w}$ (512, 2048, 8912) as function of $s^{.5}\equiv (t/t_{w})^{.5}$.}
\label{CORRS66}
\end{figure}

In fig (\ref{CORRS66}) we plot the correlation function at $\Gamma=1.8$ for $N=66$ at different 
values of $t_{w}$ (512, 2048, 8912) as function of $s^{.5}\equiv (t/t_{w})^{.5}$.  

We have plotted the data as function of $s^{.5}$ and not of $s$ for two reason:
\begin{itemize}
\item To decompress FDT region at s=0 in order to see better the building of the discontinuity at 
$s=0$
\item To expose the apparent stretched exponential decay of the correlation function for large 
values of $s$.
\end{itemize}

As we can see the two regions $FDT$ and aging are quite clear.  It is also evident that $C(\infty)$ 
is different from zero at this temperature (it is obviously zero in the high temperature phase; we 
have checked this result, but for reasons of space we do not show the corresponding data).  There is 
an overall drift of the correlation as function $t_{w}$ which seems to disappear at large times.  
This is not a surprise because we have seen that also the correlation function at $s=0$ show a 
residual dependence on $t_{w}$.

In order to see how simple aging is satisfied for other quantities we introduce the quantity 
$Q(t,t_{w})$ (introduced in \cite{PAAGE1}), defined as:
\be
Q(t,t_w) \equiv\sum_{i,k=1,N} {f(x_{i}(t+t_{w})-x_{k}(t_{w})) \over N^{2}}\label{LATER}
\ee
where we have chosen the function $f$ in an appropriate way, i.e.
\be f(x)={d^{12} \over x^{12}+d^{12}}, \ee
with $d=.3$. The function $f$ is very small when $x>>.3$ and near to $1$ for $x<.3$. 

The value of $Q$ will be a number very near to $1$ for similar configurations (in which the 
particles have moved of less than $a$) and it will be much smaller value (less than .1) for 
unrelated configurations; using the same terminology as in spin glasses \cite {EA,mpv,parisibook2} 
$Q$ can be called the overlap of the two configurations: with good approximation $Q$ counts the 
fraction of particles which have moved less that $d$.

\begin{figure}[htbp]
\epsfxsize=400pt\epsffile[22 206 549 549]{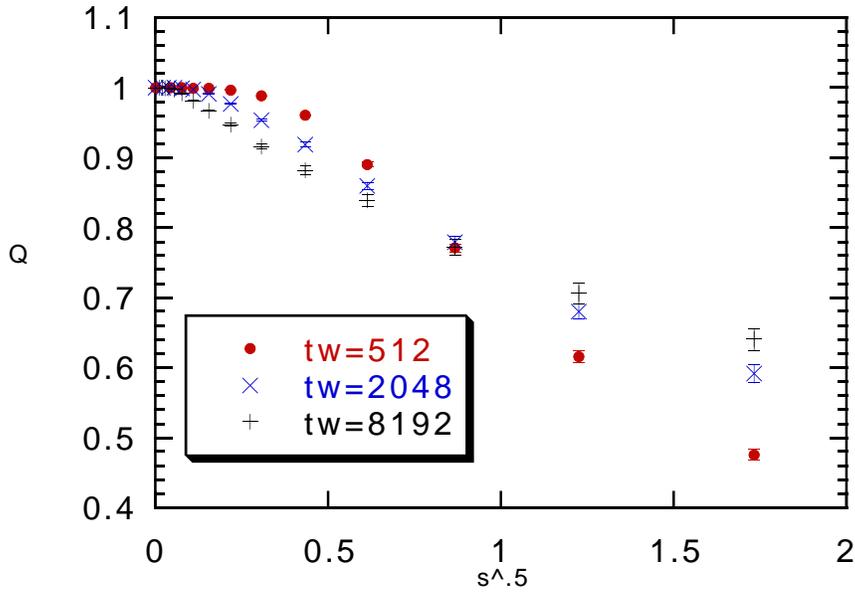}
\caption{The overlap Q at $\Gamma=1.8$ for $N=66$ at
different values of $t_{w}$ (512, 2048, 8912) as function of $s^{.5}\equiv (t/t_{w})^{.5}$.}
\label{Q66}
\end{figure}

The data for $Q$ are shown in fig.  (\ref{Q66}).  We see that also in this case we have some 
violation of simple scaling, but the violations are definitely smaller, also because, due the its 
definition the quantity $Q$ is normalized to $1$ at $s=0$.

\subsection{The response function}

We have followed a standard procedure \cite{CKR,FRARIE} to measure the off-equilibrium response 
function in simulations: we have kept the system in presence of an external field $\eps$ up to time 
$t_{w}$ and we have removed the field just at this time.

If $\eps$ is sufficient small, we have that the stress as function of time, is related to the 
integrated response $R$ by the relation
\be
{a(t+t_{w}) \over \eps} \equiv S(t+t_{w})=  R(t+t_w,0)-R(t,t_{w})\label{MISURA}
\ee
As far as the limit of $R(t,t_{w})$ when $t\to \infty$ does not depend on $t_{w}$ the quantity 
defined in eq. (\ref{MISURA}) goes to zero when $t\to \infty$.  In this way we can get the value of the 
integrated response by measuring the stress density as function of time

In our case (where we use the stress as a perturbation) the physical interpretation of the procedure 
is quite clear.  We start by putting the systems in a box which is not cubic (because $\eps\ne 0)$, 
but two sides are slightly longer of the third.  At time $t_{w}$ we change the form of the box to a 
cubic one.  In this way we deform the the system and we induce a stress which will be eventually 
decay.  In the high temperature phase, where the system is liquid, the stress will disappear in a 
short time.  On the contrary, in the glassy phase, we shall see that the stress remains for a much 
longer time (as expected
\cite{POLI}) and it shows an interesting aging behaviour.

The choice of $\eps$ is crucial.  In principle its value should be infinitesimal.  However the 
signal is proportional to $\eps$ while the errors are $\eps$ independent.  The errors on the 
response function grow as $\eps^{-1}$.  On the other hand if we take a too large value of $\eps$ we 
enter in a non linear region.  The analysis of the non linear dependence of the stress as function 
of $\eps$ would be very interesting, but it goes beyond the scope of this paper.  Here we restrict to 
the linear region.  We have taken data at $\eps=.1$ and $\eps=.05$ and we have seen that there are 
some non-linear effects.  No non-linear effects have been detected at $\eps=.02$ and $\eps=.01$.  
All the data we present in this paper come from $\eps=.01$ and they are reasonable free of systematic 
effects.  As a further check we have compared the value of $Q$ measured at $\eps=0$, as in the 
previous subsection, and the value of $Q$ at $\eps=.01$ and they differs by less than 1\%.

\begin{figure}[htbp]
\epsfxsize=400pt\epsffile[22 206 549 549]{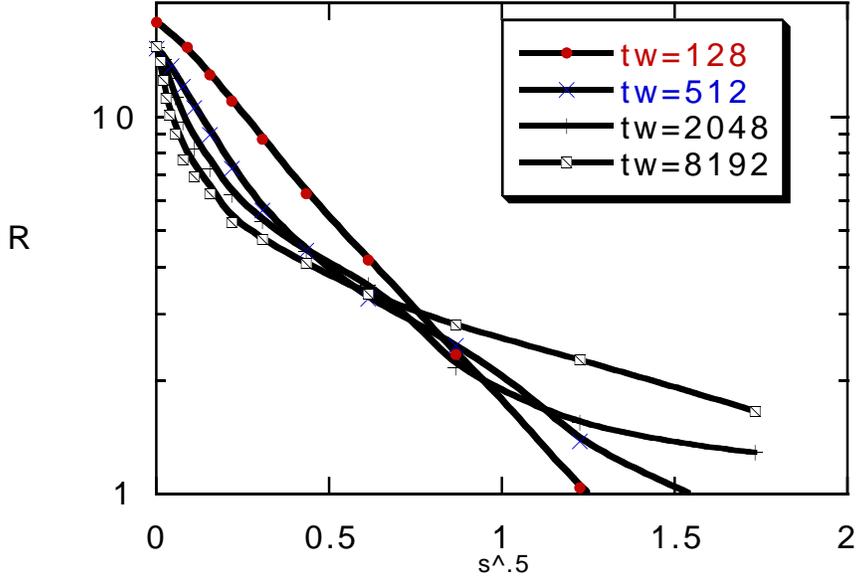}
\caption{On a logarithmic scale we see the response function at $\Gamma=1.8$ for $N=66$ at
different values of $t_{w}$ (128,512, 2048, 8192) as function of $s^{.5}\equiv (t/t_{w})^{.5}$.}
\label{RS66}
\end{figure}

In fig. (\ref{RS66}) we plot the response function $S$ at $\Gamma=1.8$ for $N=66$ at different values 
of $t_{w}$ (128, 512, 2048) as function of $s^{.5}\equiv (t/t_{w})^{.5}$.  In this case aging show 
much smaller violations than in the case of the correlations (due the fact that the value at $s=0$ 
is much less dependent on time).  Also in this case we see that the building of a discontinuity at 
$s=0$ followed by a smooth function of $s$.

\subsection{Fluctuations and response together}

The crucial step would be now to plot the fluctuations and the response together.
The previous equation tell us that
\be
{\partial S \over \partial C} = X(C)
\ee
so that it is convenient to plot $S$ versus $C$.  The slope of the function $S(C)$ is thus $X(C)$.

\begin{figure}[htbp]
\epsfxsize=400pt\epsffile[22 206 549 549]{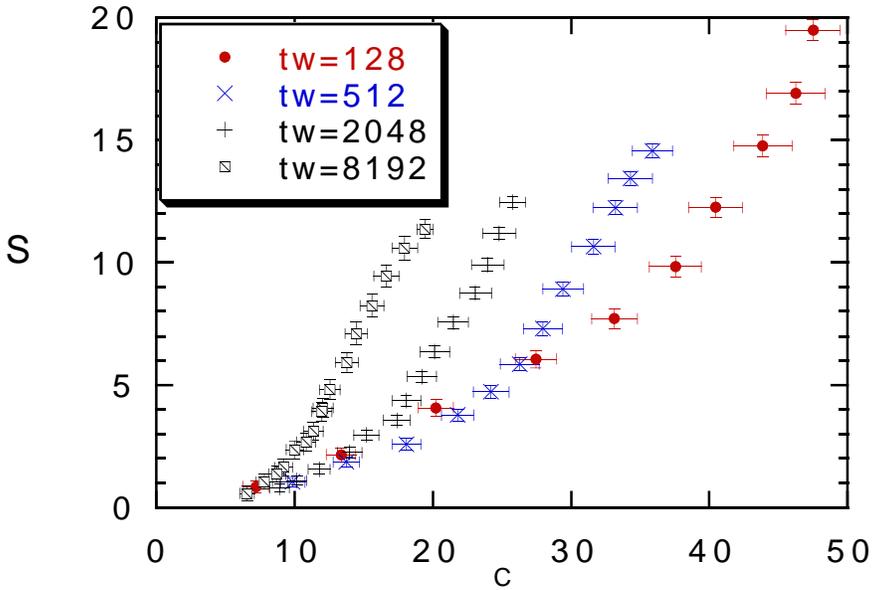}
\caption{The response $S$ as function of $C$ at  $\Gamma=1.8$ for $N=34 $ at
different values of $t_{w}$ (128, 512, 2048, 8912).}
\label{CK34}
\end{figure}

 In the one step replica symmetry breaking scheme we expect that:
\bea
S= mC \for C< C(\infty),\\
S= C+ (1-m) C(\infty) \for C< C(\infty).
\eea

In fig. (\ref{CK34}) we find the data for the response $S$ as function of $C$ at $\Gamma=1.8$ for 
$N=34 $ at different values of $t_{w}$ (128, 512, 2048, 8912).  We can see that also here there is 
strong dependance on $t_{w}$.  We can distinguish two regions: 
\begin{itemize}
\item
A high $C$ region (typically $C>C(\infty)$), where the function $S$ has nearly slope one.
\item 
A region of smaller $C$, where the dependence of $S$ on $C$ is more complex.
\end{itemize}
We remark that the data for smaller value of the waiting time seems to be very similar to the 
predictions of the one step replica theory but this conclusion is not so clear for larger values of 
the waiting time.

\begin{figure}[htbp]
\epsfxsize=400pt\epsffile[22 206 549 549]{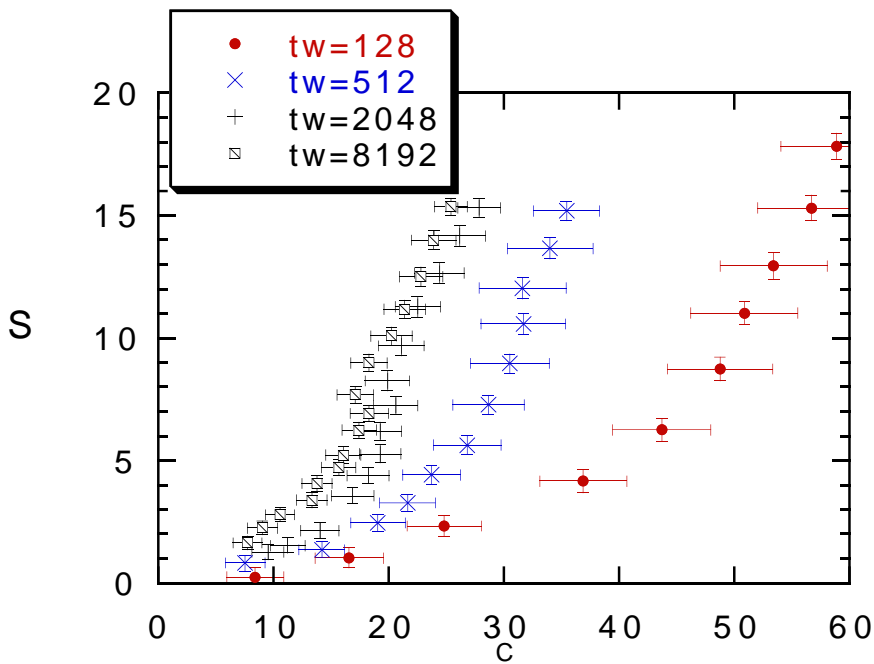}
\caption{The response $S$ as function of $C$ at  $\Gamma=1.8$ for $N=66 $ at
different values of $t_{w}$ (128 512, 2048, 8192).}
\label{CK66}
\end{figure}

In fig.  (\ref{CK66}) we find the data for the response $S$ as function of $C$ at $\Gamma=1.8$ for 
$N=66 $ at different values of $t_{w}$ (128, 512, 2048, 8192).  We find a behaviour qualitatively 
similar to the one with a smaller number of particles at small waiting time; however here the data 
at $t_{w}=2048$ and $t_{w}=8192$ are much more similar, giving us the hope to be near to the 
asymptotic limit in the case of this larger system.

\begin{figure}[htbp]
\epsfxsize=400pt\epsffile[22 206 549 549]{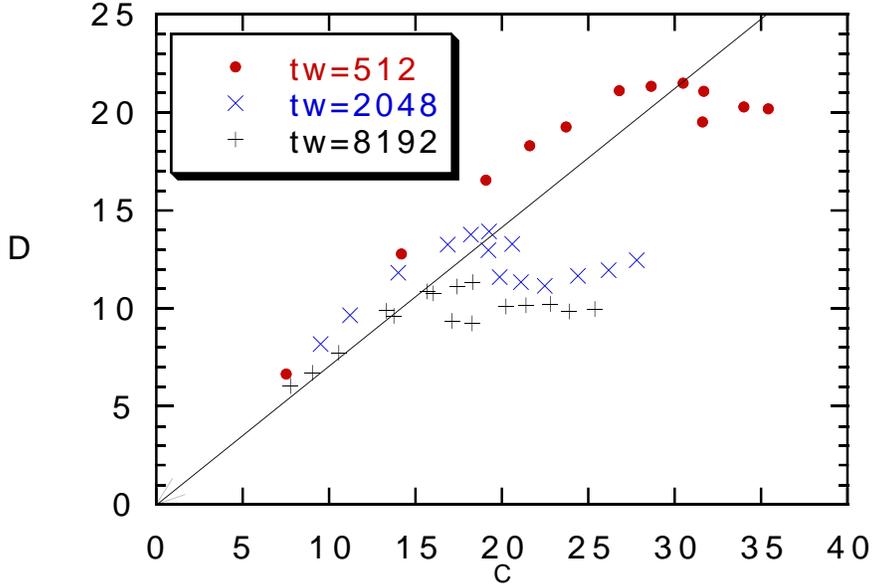}
\caption{The difference $D\equiv C-S$ as function of $C$ at $\Gamma=1.8$ for $N=34 $ at
different values of $t_{w}$ (512, 2048, 8912).  The straight line correspond to one step replica 
symmetry breaking with $m=.3$.}
\label{D34}
\end{figure}

In order to expose the existence of a region where the FTD is valid \cite{FRARIE}, it is convenient 
to define the function $D\equiv C-S$.  It is evident that in the FDT region the function $D(C)$ must 
be equal to a constant.  In order to test this prediction and to verify the existence of a region 
where the FDT relation holds, in fig.  (\ref{D34}) we plot the the function $D$ versus $S$ at 
$\Gamma=1.8$ for $N=66 $ at different values of $t_{w}$ (512, 2048, 8912).  A plateaux region is 
quite evident.  The level of the plateaux is still dependent on $t_{w}$.  The straight line 
correspond to one step replica symmetry breaking with $m=.3$.

The behaviour of the function $S(C)$ at small $C$ would be quite interesting to assess.  We have 
three possibilities of increasing complexity
\begin{itemize}
\item The function $S(C)$ is zero for $C<C(\infty)$.  This correspond to the simple situation where
replica symmetry is not broken.
\item The function $S(C)$ is linear for $C<C(\infty)$ with slope $m$.  This correspond to one step
replica symmetry breaking.
\item The function $S(C)$ has a power behaviour for $C<C(\infty)$ with exponent greater than one.  
In this case the replica symmetry is broken in a continuous way.
\end{itemize}

\begin{figure}[htbp]
\epsfxsize=400pt\epsffile[22 206 549 549]{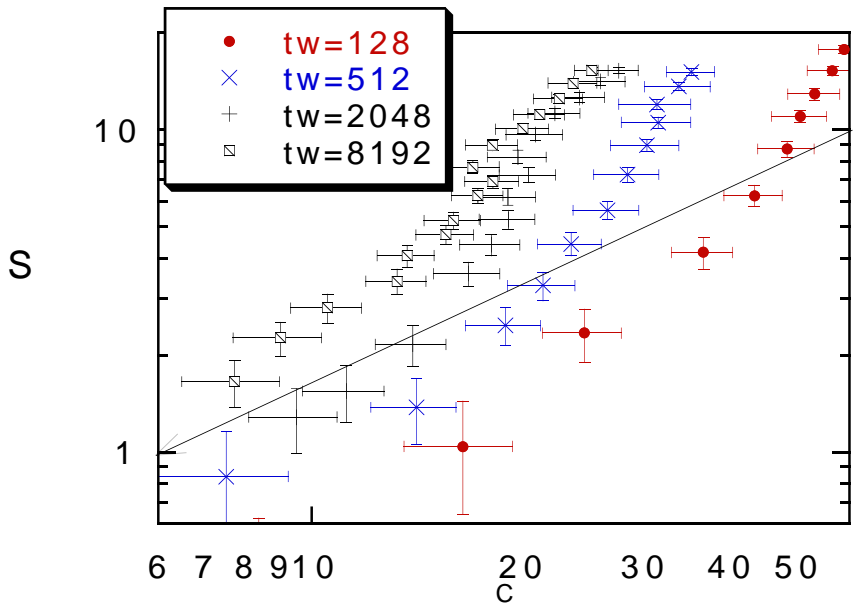}
\caption{The response $S$ as function of $C$ at  $\Gamma=1.8$ for $N=66 $ at
different values of $t_{w}$ (128 512, 2048, 8192) on a double logarithmic scale.  The straight line 
correspond to a linear behaviour.}
\label{CKLN66}
\end{figure}

The previous data for $S(C)$ plotted in double logarithmic scale may be useful to clarify this point, 
see fig.  (\ref{CKLN66}).  We find that the data for small $t_{w}$ seem to be linear in agreement 
with one step replica symmetry breaking predictions.  There is a rather strong dependence on $t_{w}$ 
in this region and the extrapolation to $t_{w}=\infty$ cannot safely done.  A value of $m$ around 
$.3$ (or more) is compatible with the data.  This is an interesting problem that must be investigated 
further.

\section{Conclusions}

It has been shown in the previous section that there is a low temperature region where the large 
time extrapolation of the energy and of the stress autocorrelation function can be done by assuming  
simple power low corrections.  The extrapolated value quite likely do not correspond to equilibrium 
values, but to metastable values and the real equilibrium values are reached only at much bigger 
time.

The behaviour in this region is quite different from the high temperature region.  The 
autocorrelation function of the stress is bigger than the analytic continuation of its value from 
the high temperature region.
Approximate simple aging is found.  The function $X(C)$ of the CK theory has been computed.  Also 
this function shows a dependence on the value of $t_{w}$ which seems to decrease by increasing 
$t_{w}$.  The extrapolation of the function $X(C)$ at infinite (or very large) $t_{w}$ it is a 
delicate problem, that should be treated in a careful way.  At first sight is seems likely that in 
order to get a reliable extrapolation one need to increase the time $t_{w}$ by a factor 10, increase 
somewhat the statistics in order to decrease the errors and simultaneously go to larger samples in 
order to avoid finite volume effects, which become stronger and stronger when the time increase 
\footnote{We should also increase the value of $s$ in order to explore better the region where $C$
becomes small.}.  One probably needs a factor 100 more in computer time, which is not a very large 
amount (considered that this computation has taken a few weeks of a workstation), but it would 
certain go outside the exploratory aim of this work and of the capabilities of the hardware we have 
used.

It is likely that something smarter can be done.  Two possibilities are immediate:
\begin{itemize}
\item We can repeat the same procedure where the temperature is slowed cooled in steps of total 
length $t_{w}$ starting from a finite temperature to the final temperature.  In this way is possible 
that aging and the other phenomena survive, but the finite time corrections could be much smaller
\cite{MAPAZU}.
\item We could change the quantity we study.  It may be possible that the stress is not the best 
suited quantity to investigate; indeed the quantity $Q$ seems to satisfy much better simple aging 
with much smaller corrections.
\end{itemize}

It seems to me that there is a quite large amount of phenomena to investigate in the approach to 
equilibrium of glasses.  It would be interesting to see if the indications given in this paper will 
be confirmed by more accurate and lengthy computations.

\section* {Acknowledgments} I thank S.  Franz and D.  Lancaster for useful discussions and L.
Cugliandolo and J. Kurchan for having suggested to me to measure of the stress after changing the 
form of the box.

\section*{ Appendix: spin glasses}

In this appendix I will briefly recall the situation for spin glasses.
\begin{figure}[htbp]
	\epsfxsize=400pt\epsffile[22 206 549 549]{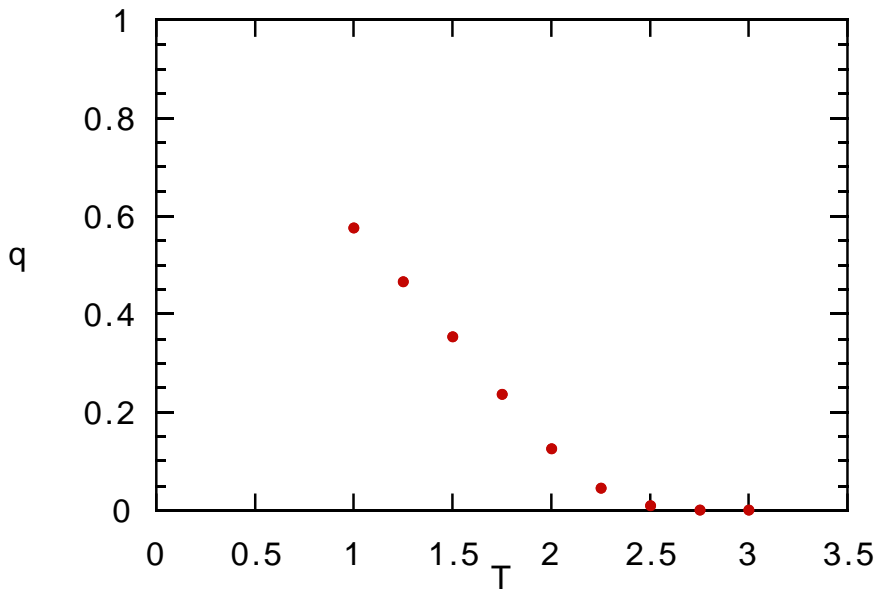}
\caption{The value of $q$ as function of the temperature in a 4-dimensional spin glass.}   
\label{Q}
\end{figure}
Aging and the predictions of the CK theory have been carefully analyzed in 
\cite{FRARIE}: a comparison of their results with our would be instructive.  Here we limit ourself 
to the following general remark.

 In spin glasses the relevant quantity is the total magnetization
\be
M=\sum_{i}\si_{i}.
\ee
It is easy to see that at zero magnetic field in the random bond model (due to gauge invariance) the 
following relation holds at all temperatures.
\be
{ <M(t)>^{2} \over N} =1.
\ee

We can now compute in simulations the magnetic susceptibility ($\chi$), i.e.  by inserting a small 
magnetic field $h$ and my measuring the ratio $<M>/h$.  Following the previous discussion (\ref{QA}) 
we have that
\be
\chi=\beta(1-q).
\ee

An example of the function $q$ computed in simulations for spin glasses (a four dimensional model) is 
showed in fig.  (\ref{Q}), where the data are taken from \cite{MAPAZU}. 

There is a striking difference from spin glasses and the case of binary glasses presented here, 
which it is worthwhile to note:
\begin{itemize}
\item In glasses the response to the stress tensor is computed from symmetry arguments and the 
transition is present in the fluctuations.
\item 
In glasses the fluctuations of the magnetization are computed from symmetry arguments and the 
transition is present response, (i.e.  the susceptibility).
\end{itemize}

However it seems to me that this difference does not seem to have deep physical consequence and it 
is just an effect of the different choice of observables.

\end{document}